\documentclass[aps,reprint]{revtex4-1}
\usepackage{amsmath,amssymb,bm}
\usepackage{balance}
\usepackage{graphicx}
\usepackage{epstopdf}
\usepackage{natbib}

\newcommand{\pe}{\text{Pe}}
\newcommand{\blu}[1]{#1}

\begin{document}

\title{Orientational instability and spontaneous rotation of active nematic droplets}
\author{Matvey Morozov}
\author{S{\'e}bastien Michelin}
\email{sebastien.michelin@ladhyx.polytechnique.fr}
\affiliation{
  LadHyX -- D{\'e}partement de M{\'e}canique, CNRS -- {\'E}cole Polytechnique, 
  Institut Polytechnique de Paris, 
  91128 Palaiseau Cedex, France}

\begin{abstract}
In experiments, an individual chemically active liquid crystal (LC) droplet submerged in the bulk of a  surfactant solution may self-propel along a straight, helical, or random trajectory. In this paper, we develop a minimal model capturing all three types of self-propulsion trajectories of a drop in the case of a nematic LC with homeotropic anchoring at LC-fluid interface. We emulate the director field within the drop by a single preferred polarization vector that is subject of two reorientation mechanisms, namely, the internal flow-induced displacement of the hedgehog defect and the droplet's rotation. Within this reduced-order model, the coupling between the nematic ordering of the drop and the surfactant transport is represented by variations of the droplet's interfacial properties with nematic polarization. Our analysis reveals that a novel mode of orientational instability emerges from the competition of the two reorientation mechanisms and is characterized by a spontaneous rotation of the self-propelling drop responsible for helical self-propulsion trajectories. In turn, we also show that random trajectories in isotropic and nematic drops alike stem from the advection-driven transition to chaos. The succession of the different propulsion modes is consistent with experimentally-reported transitions in the shape of droplet trajectories as the drop size is varied.
\end{abstract}

\maketitle

\section{Introduction}
Numerous life forms around us feature left-right asymmetry or chirality, that is not externally imposed but arises spontaneously through symmetry-breaking events, as in the course of embryogenesis~\cite{Levin05, MartinDuran16}. There exist experimental evidence that in morphogenesis chirality emerges at the level of individual cells and impacts all subsequent stages of the process~\cite{Taniguchi11, Chen12, Inaki18}. Chirality of individual cells also affects their mechanical characteristics: individual epithelial cells suspended in the bulk of protein solution may start rotating spontaneously and the direction of this rotational motion is linked to the chirality of the cell~\cite{Chin18}. Beyond the biological realm, spontaneous emergence of asymmetry has been identified at the individual or collective level in the dynamics of chemically-active colloids and droplets~\cite{Thutupalli11,Izri14,Nagasaka17, Zwicker17}. Here, we demonstrate that chirality may similarly arise spontaneously in isotropic active colloidal systems using a chemically-active nematic droplet as a model object.

The behavior of liquid crystal (LC) droplets suspended in a bulk fluid is determined by the interplay between hydrodynamic stresses, interfacial tension, surface anchoring and liquid crystal elasticity~\cite{Lavrentovich98, Rey99}. For instance, the director field configuration within an LC drop may be altered by a change in the bulk fluid's chemical composition~\cite{Tran17}. It is also possible to transform the structure of the droplet interface by tuning its LC configuration~\cite{Wang14}. In the examples above, the behavior of LC drops is controlled externally. In contrast, active droplets may act autonomously, powered by the energy released through their chemical activity~\cite{Herminghaus14, Maass16}. In particular, nematic droplets undergoing gradual solubilization in aqueous surfactant solutions may self-propel spontaneously, while their smectic counterparts exhibit formation of filament-like structures~\cite{Peddireddy12}. Chirality of cholesteric active drops allows them to self-propel along a helical trajectory~\cite{Yamamoto17}.

We focus here on the behavior of active nematic drops with homeotropic anchoring that are completely isotropic in the absence of any hydrodynamic flow or droplet motion. Such active nematic microdroplets self-propelling in the bulk of surfactant solution were recently observed to alter their propulsion regime depending on the ordering in LC phase: drops in nematic state exhibit helical self-propulsion trajectories that were not found in isotropic droplets~\cite{Kruger16, Suga18}. Emergence of the curly trajectories is particularly intriguing, since nematic droplets possess no intrinsic chirality. In that regard, the nature of spontaneous curling of self-propulsion trajectories in nematic drops is fundamentally different from the similar phenomenon observed in cholesteric drops~\cite{Yamamoto17}. It was further reported that in experiments the hedgehog defect located in the center of a motionless nematic drop with homeotropic anchoring is displaced and becomes a boojum when the droplet self-propels, yet another manifestation of the intimate coupling of the hydrodynamic stresses and elasticity of the LC phase~\cite{Kruger16}. This observation led to the assumption that the helical motion results from anisotropic stresses caused by the defect displacement which, in turn, is sustained by the flow in the LC. However, a detailed modeling of the coupling of hydrodynamics, physico-chemical activity and transport, and internal droplet structure and elasticity is still lacking, which has so far prevented a full understanding of the emergence of spontaneous chirality in the droplet trajectories.

In addition to straight and helical self-propulsion trajectories, active nematic drops also exhibit chaotic behavior~\cite{Suga18}. Chaotic self-propulsion trajectories were also observed in isotropic droplets~\cite{Izri14, Moerman17, Suga18}, suggesting the existence of a universal mechanism of transition to chaos in active drops. It was recently argued that emergence of random behavior of isotropic active drops is linked to the nonlinear effect of surfactant advection and to the precise physico-chemical mechanism responsible for the flow actuation~\cite{Morozov19b}. Specifically, the actuation mechanisms of the flow from local physico-chemical gradients along a fluid-fluid interface can be decomposed into two categories: (i) phoretic phenomena typically modeled by a discontinuity of the flow velocity across the droplet's surface and (ii) the Marangoni effect associated with discontinuous stresses at the interface~\cite{Anderson89}. The latter is typically considered as the sole mechanism of active droplet mobility~\cite{Herminghaus14, Maass16}. Yet, recent numerical simulations indicate that, in the case when the mobility mechanisms from the two categories act simultaneously, the Marangoni effect may hinder the transition to chaos~\cite{Morozov19b}. On the other hand, ubiquity of chaotic regimes in experiments with active droplets hints that some contribution of diffusiophoresis might be important for successful modeling of surfactant-laden interfaces of active LC droplets~\cite{Izri14, Moerman17, Suga18}.

In this paper, we propose a minimal model of an active nematic drop suspended in a surfactant solution. Our model captures all three types of self-propulsion trajectories observed experimentally in active nematic drops of increasing radius, i.e. straight, helical, and random trajectories. We employ our model to explain the emergence of chiral trajectories in nematic drops and elucidate the transition to chaos for drops in both isotropic and nematic states. The paper is organized as follows. The problem, model equations and relevant dimensionless parameters are introduced in Sec.~\ref{problem}, while Sec.~\ref{results} presents the results of 3D numerical simulations of an active drop in isotropic and nematic states. Finally, we discuss our findings and perspectives opened by this work in Sec.~\ref{discussion}.

\section{Physical model of a nematic active droplet}
\label{problem}
A full modeling of the nematic ordering of the droplet as a result of its coupling to the internal and external flows and surfactant concentration is particularly challenging~\cite{Lavrentovich98, Lin14, Kitavtsev18}. Thus, our primary goal here is to provide a generic physical insight on the critical role of the coupling between these phenomena in setting the dynamics and enabling different types of trajectories for active LC microdroplets.

\begin{figure}
  \centering
  \includegraphics[scale=0.55]{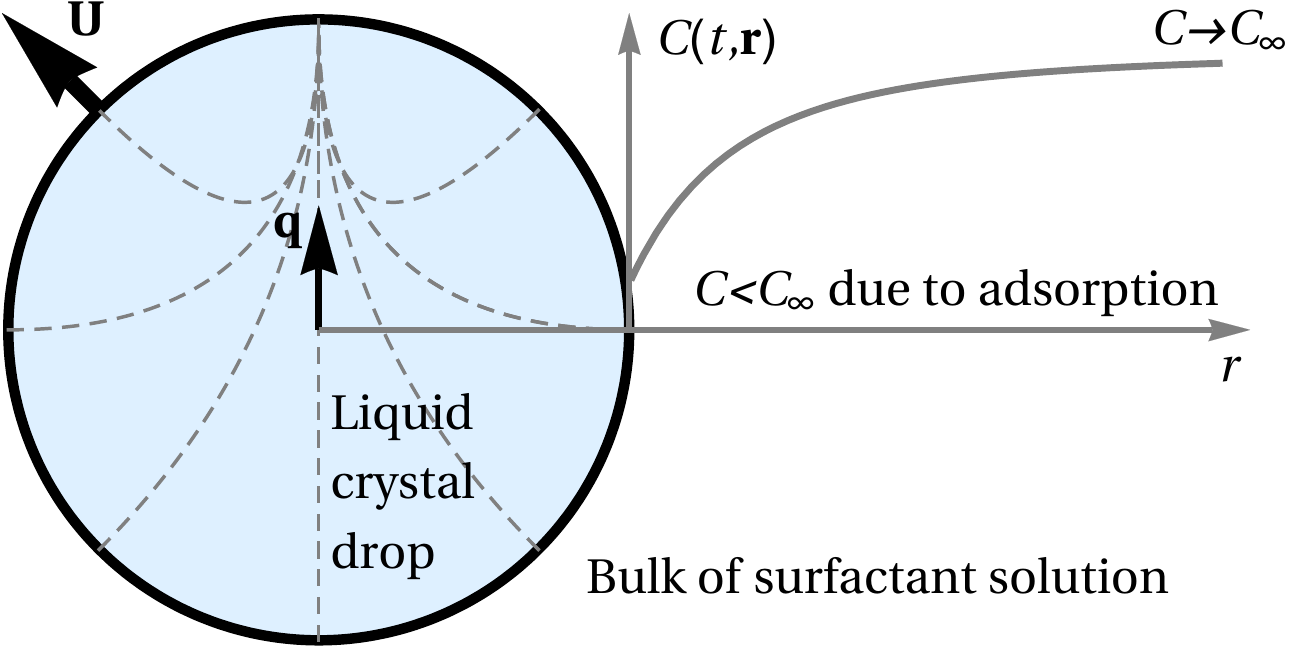}
  \caption{
    Model: an active liquid crystal (LC) drop moving with velocity $\textbf{U}$ through the bulk of surfactant solution. The director field in the drop (dashed gray lines) is emulated by a single vector $\textbf{q}$ evolving according to Eq.~\eqref{displacement}. The continual surfactant adsorption generates concentration gradients around the drop that result in phoretic slip, Eq.~\eqref{slip}, and self-propulsion. Note that $\textbf{U}$ and $\textbf{q}$ are not necessarily aligned due to the unsteady internal dynamics of the nematic field, Eq.~\eqref{displacement}.
  }
  \label{fig0}
\end{figure}

Accordingly, we do not describe the director field of the LC structure explicitly, but retain its two-way coupling to the hydrodynamic problem by characterizing the \blu{cumulative} polarization of the director field \blu{within the droplet} by a single vector $\textbf{q}(t)$ that emulates the preferred orientation of nematic molecules in the drop as sketched in Fig.~\ref{fig0}. Naturally, ${\textbf{q}=\textbf{0}}$ for a motionless droplet with homeotropic anchoring. When the droplet self-propels with a constant translation velocity $\mathbf{U}$ and no rotation ($\boldsymbol\Omega=0$), the flow and director fields are axisymmetric~\cite{Kruger16} and the polarization of the droplet LC structure is increased for higher velocity. As a result, we postulate that ${\textbf{q} = {\cal K} \textbf{U}}$ in such a case, where ${{\cal K} > 0}$ is a constant coefficient quantifying the effect of the flow on the nematic ordering within the drop. At last, we also assume that $\textbf{q}$ rotates with the droplet, namely,
\begin{equation}
  \label{displacement}
  \partial_t \textbf{q} 
    = -{\cal B} \left( \textbf{q} - {\cal K} \textbf{U} \right)
      + \boldsymbol{\Omega} \times \textbf{q},
\end{equation}
where ${\cal B}^{-1}$ is the characteristic relaxation time required for realignment of $\textbf{q}$ along $\textbf{U}$. Equation~\eqref{displacement} is used in the following as a minimal coupling model of the nematic ordering within the drop with the flow field.

To capture transition to chaos in our model while retaining the simplest framework, we consider only phoretic mobility (i.e. the Marangoni effect is neglected); further, we assume that the inner fluid is much more viscous than the bulk phase.\footnote[2]{Note that the model could be straightforwardly generalized to include both Marangoni and phoretic effects as well as arbitrary viscosity ratios.~\cite{Morozov19b}
} As a result, assuming the droplet remains spherical at all times, the outer flow velocity at the droplet interface reads (in the reference frame translating with the droplet),
\begin{equation}
  \label{slip}
  \textbf{u} = \boldsymbol{\Omega} \times \textbf{r}
    - M\left( \textbf{q}, \textbf{r} \right) \nabla_s C
  \quad \text{at  } r = R,
\end{equation}
where $C(t,\textbf{r})$ denotes surfactant concentration, ${M\left( \textbf{q}, \textbf{r} \right)}$ is the phoretic mobility coefficient, $R$ is the droplet radius, and ${\nabla_s \equiv \left( \mathbf{I} - \textbf{n}\textbf{n} \right) \cdot \nabla}$ with $\mathbf{I}$ and $\textbf{n}$ denoting the identity tensor and the outward normal to the droplet surface, respectively. In contrast to earlier works~\cite{Michelin13b, Morozov19b}, the phoretic mobility, ${M\left( \textbf{q}, \textbf{r} \right)}$, is not a constant in Eq.~\eqref{slip}. Instead, its definition accounts for the anisotropy of droplet's interfacial properties associated with the spontaneous polarization of the LC phase, $\mathbf{q}$,
\begin{equation}
  \label{mobility}
  M\left( \textbf{q}, \textbf{r} \right)
    = M_0 + M_1 \textbf{q} \cdot \textbf{r},
\end{equation}
where ${M_0 > 0}$ and $M_1$ are isotropic and anisotropic mobility coefficients, respectively. Together Eqs.~\eqref{displacement} and~\eqref{slip} constitute a minimal model of coupling between the flow field around the drop and the director field in the inner LC phase. We emphasize that active droplet mobility is typically attributed solely to the Marangoni effect~\cite{Herminghaus14, Maass16}; Equation~\eqref{slip} thus represents an alternative approach to modeling of active drop mobility and the secondary goal of this paper is to demonstrate that such a model can successfully capture fundamental dynamical features of active drops, including the onset of chaos.

In experiments, the energy required for self-propulsion of active drops is generated in a chemical reaction sustained at the droplet interface~\cite{Izri14, Kruger16, Moerman17, Suga18}. Specifically, drops undergo gradual micellar dissolution and their dissolution time is orders of magnitude larger than the characteristic time scales associated with their propulsion. Accordingly, we treat their volume and the dissolution process as quasi-static. We further assume that nematic ordering does not impact dissolution intensity so that reaction at the droplet surface has a fixed rate ${\cal A} > 0$,
\begin{equation}
  \label{c_cond}
  {\cal D} \partial_r C = {\cal A}
  \quad \text{at  } r = R.
\end{equation}

Inertial effects on the flow stirred by a microscopic droplet are negligible, and the outer flow field thus satisfies Stokes' equations,
\begin{equation}
  \label{eqs_flow}
  \nabla \cdot \textbf{u} = 0, \quad
  \nabla P = \eta \nabla^2 \textbf{u},
\end{equation}
where $\eta$ is the bulk fluid viscosity. Unlike inertia, surfactant advection around the drop is not negligible~\cite{Michelin11, Michelin13b} and surfactant transport around the drop is governed by an advection-diffusion equation,
\begin{equation}
  \label{eqs_ad}
  \partial_t C + \textbf{u} \cdot \nabla C = {\cal D} \nabla^2 C,
\end{equation}
with ${\cal D}$ the surfactant diffusivity in the outer phase.

Finally, away from the active drop, flow velocity and surfactant concentration attain constant values,
\begin{equation}
  \label{far}
  \textbf{u} = -\textbf{U}, \quad 
  C = {\cal C}_\infty,
\end{equation}
and the translation and rotation velocities, $\textbf{U}$ and $\boldsymbol{\Omega}$, are finally determined by enforcing that the total hydrodynamic force and torque on the drop must vanish.

In the following the problem is non-dimensionalized by choosing $R$,  ${R / {\cal V}}$ and ${{\cal A} R / {\cal D}}$  as characteristic scales for length, time and relative surfactant concentration (i.e. $C-\mathcal{C}_\infty$), respectively. Here, the characteristic velocity scale ${\cal V}$ is chosen as that of an inert phoretic particle within a uniform concentration gradient ${{\cal A} / {\cal D}}$, namely, ${{\cal V} \equiv {\cal A} M_0 / {\cal D}}$~\cite{Anderson89}.

Note that our model is based on an arbitrary definition of the dimensional magnitude of $\textbf{q}$. After scaling $\textbf{q}$ with ${{\cal K} {\cal V}}$, the dimensionless form of Eqs.~\eqref{displacement}-\eqref{far}  includes only three dimensionless parameters. The first is the P{\'e}clet number, ${\pe \equiv {\cal V} R / {\cal D}}$, i.e.~the relative efficiency of advective transport and diffusion of surfactant around the drop. The second is the polarizability coefficient, ${m_1 \equiv {\cal K} {\cal V} M_1 R / M_0}$, and measures how much nematic ordering changes the drop's interfacial properties. The third is the retardation ratio, ${\beta \equiv {\cal B} R / {\cal V}}$, i.e.~the ratio of the hydrodynamic time scale to the relaxation time of the nematic ordering.

\section{Results}
\label{results}
\begin{figure}
  \centering
  \includegraphics[scale=0.52]{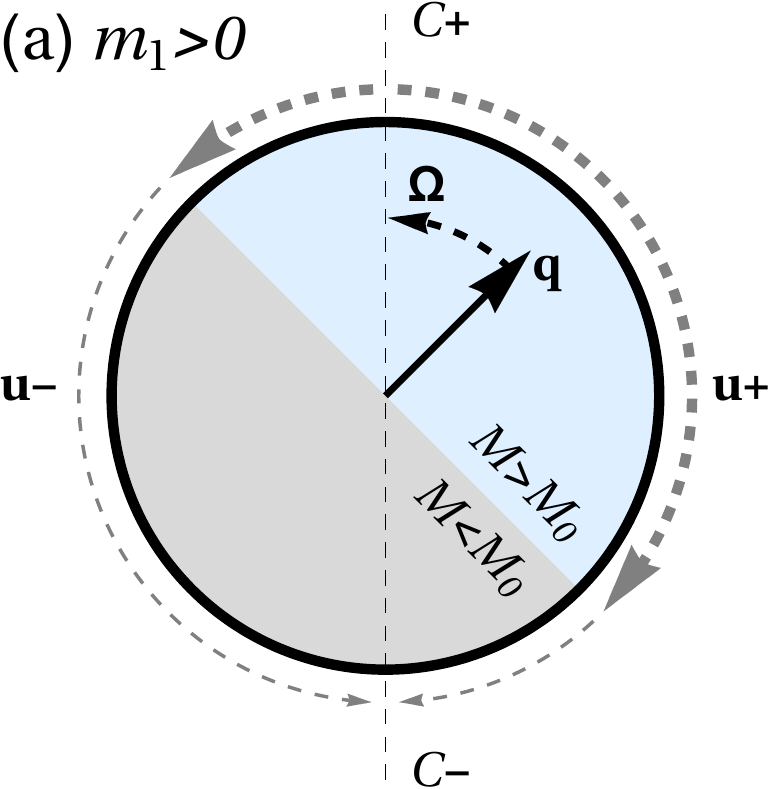}
  \includegraphics[scale=0.52]{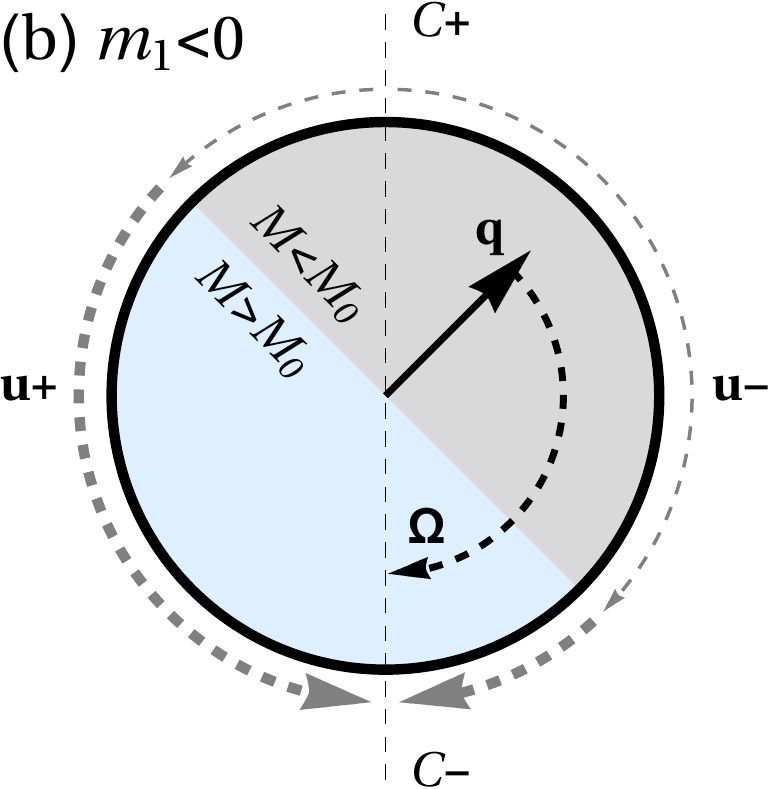}
  \caption{
    Physical interpretation of the
    rotation velocity $\boldsymbol \Omega$ of a nematic drop 
    with polarization $\textbf{q}$
    in response to a non-uniform surfactant distribution, Eq.~\eqref{omega} ($C+ > C-$).
    Anisotropic mobility $M$ at the droplet interface
    results in uneven slip velocity $\textbf{u}$
    (thicker gray arrows correspond to higher velocity).
    (a) For ${m_1 > 0}$, the drop rotates to align
    $\textbf{q}$ and the gradient of $C$.
    (b) When ${m_1 < 0}$, 
    $\textbf{q}$ is rotated away from the gradient of $C$.
  }
  \label{fig1}
\end{figure}

The reciprocal theorem for Stokes flows provides the rotation velocity of a spherical particle as a functional of the flow velocity at the particle surface~\cite{Stone96, Pak14}, and using Eq.~\eqref{slip}, leads to
\begin{equation}
  \label{omega}
  \boldsymbol \Omega = \frac{3 m_1}{2} \textbf{q} \times
    \left< \textbf{n} \left. C \right|_{r=1} \right>,
\end{equation}
where angle brackets represent surface average. The gist of Eq.~\eqref{omega} is that for ${m_1 > 0}$ (resp. $m_1<0$) the drop in nematic state rotates to align $\textbf{q}$ with (resp. against) the concentration gradient (Figure~\ref{fig1}). The case of ${m_1 < 0}$ is thus characterized by a competition of the two reorientation mechanisms of $\textbf{q}$ in Eq.~\eqref{displacement}. On the one hand, active drop self-propulsion velocity $\textbf{U}$ is typically aligned with the self-imposed concentration gradient~\cite{Michelin13b, Morozov19a, Morozov19b} and the relaxation term in Eq.~\eqref{displacement} strives to align $\textbf{q}$ and $\textbf{U}$. On the other hand, for ${m_1 < 0}$, equation~\eqref{omega} states that $\textbf{q}$ is rotated away from the concentration gradient and, consequently, away from $\textbf{U}$. Below we demonstrate that this competition enables a novel mode of instability that results in helical motion of active LC drops.

Interestingly, Eq.~\eqref{omega} also demonstrates that an isotropic droplet (${m_1 = 0}$) can not exhibit any rotation. Indeed, in the case of ${m_1 = 0}$, Eq.~\eqref{slip} establishes a linear relation of the slip velocity with the gradient of a scalar field, ${C(t, |\textbf{r}| = 1)}$, and the latter is naturally periodic for a spherical drop.

In what follows, the dimensionless form of Eqs.~\eqref{displacement}-\eqref{far} is solved numerically using spectral expansions of the flow and concentration fields (see Appendix~\ref{method}).

\subsection{Axisymmetric flow regimes}
We first focus on time-independent axisymmetric solutions corresponding to a drop self-propelling steadily along a straight line. To obtain these steady states efficiently, we employ a continuation procedure: we solve the axisymmetric evolution equations for $\textbf{q}$ and $C$ numerically in the case of ${m_1 = -2, 0, 2}$, and sequentially increasing the P{\'e}clet number, ${\pe \in [ 3.5, 27 ]}$, initializing each computation from the limit regime achieved at the previous value of $\pe$. Note from Eq.~\eqref{displacement}, that the value of the relaxation coefficient $\beta$ does not affect the steady state.
\begin{figure}
  \centering
  \includegraphics[scale=0.5]{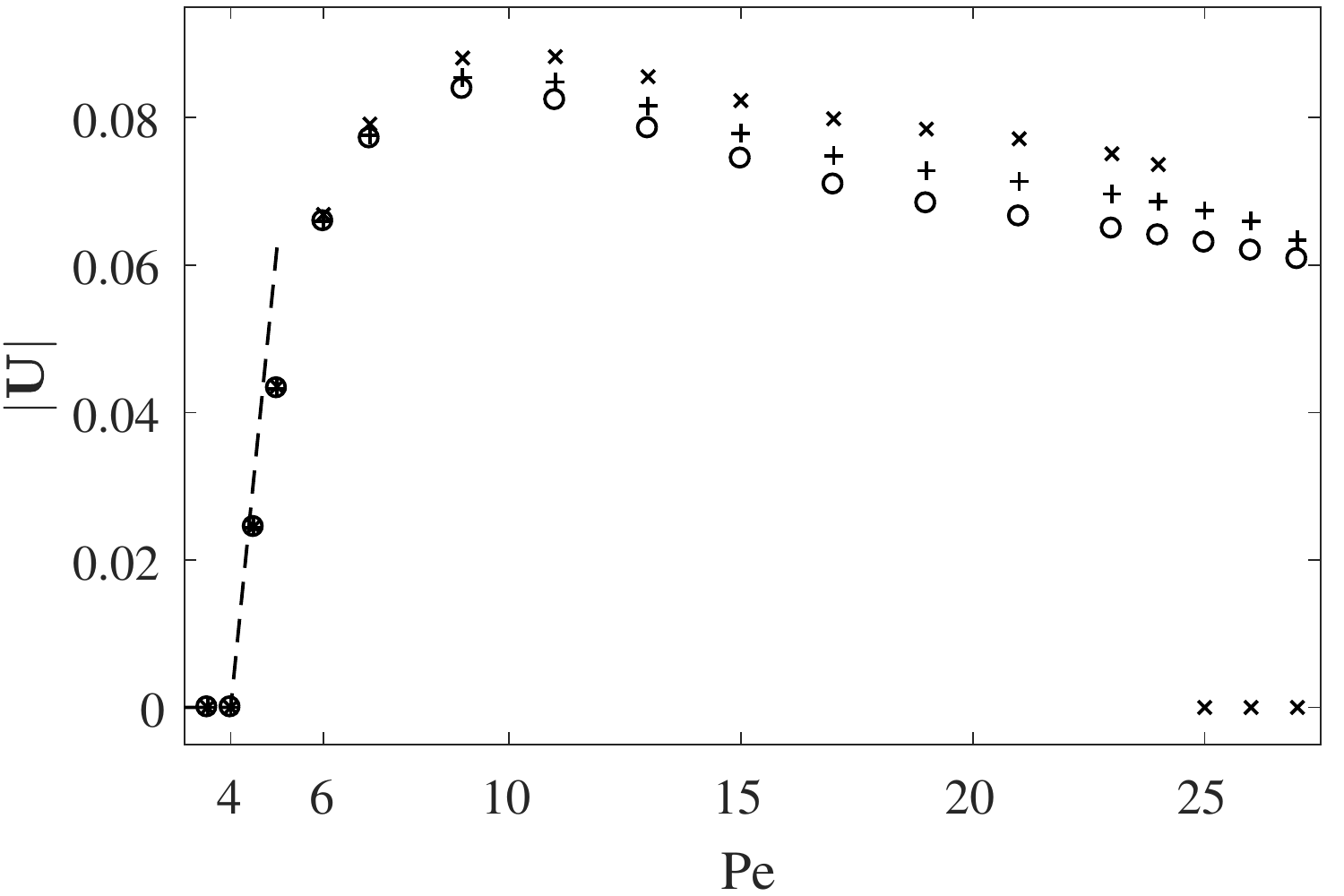}
  \caption{
    Evolution of the axisymmetric self-propulsion velocity $\textbf{U}$ of an active LC drop
    with $\pe$ for ${m_1 = 0}$ ($+$), 
    ${m_1 = 2}$ ($\times$), 
    or ${m_1 = -2}$ ($\circ$).
    The asymptotic prediction 
    for an isotropic drop (${m_1 = 0}$) in Ref.~\cite{Morozov19b} is also reported (dashed).
  }
  \label{fig3}
\end{figure}

\begin{figure}
  \centering
  \includegraphics[scale=0.55]{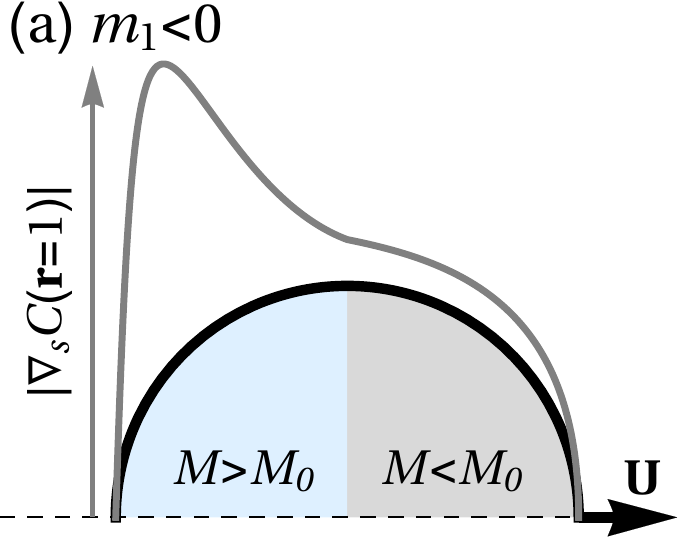}
  \includegraphics[scale=0.55]{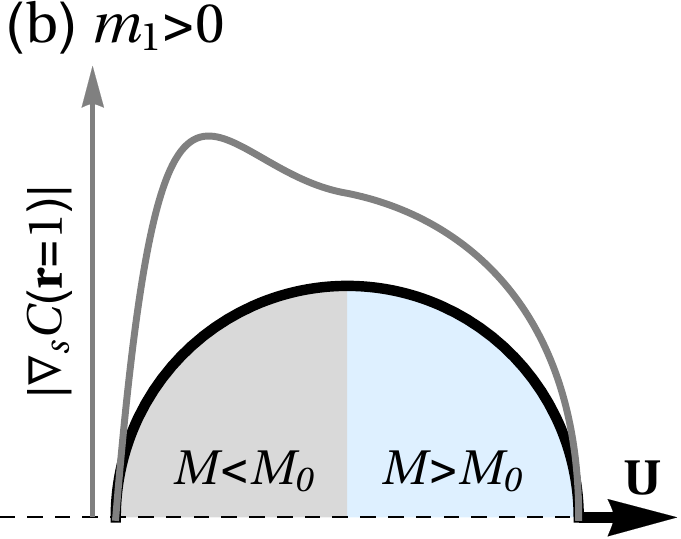}
  \caption{
    Effect of anisotropic mobility on the surfactant concentration gradient at the interface of a LC drop in steady axisymmetric self-propulsion. (a) For ${m_1 < 0}$, mobility and surfactant transport are enhanced at the back of the drop, resulting in a sharp peak in concentration gradient near the trailing pole of the drop. (b) The peak is blunted when ${m_1 > 0}$, resulting in more homogeneously distributed concentration gradients. The effect of anisotropic mobility is purposely exaggerated for illustration purpose.
  }
  \label{fig4}
\end{figure}

The results are shown on Fig.~\ref{fig3} and indicate that the self-propulsion velocity of an active LC drop depends on its polarizability: droplets with ${m_1 > 0}$ propel faster than their isotropic counterparts, while the opposite is true for ${m_1 < 0}$. We explain this effect as follows. For droplets with positive polarizability (i.e., ${m_1 > 0}$), the mobility is enhanced (resp. reduced) near the front (resp. back) pole of the propelling drop. Stronger phoretic flows at the front result in a higher supply of surfactant-rich fluid to the equatorial region of the surface, while a reduced advective transport near the back allows for a more efficient adsorption of the supplied surfactant. Consequently, the surfactant gradient and associated slip flow near the equatorial plane (that has the largest hydrodynamic influence on propulsion) is increased (Figure~\ref{fig4}), resulting in larger propulsion velocities for ${m_1 > 0}$. The same argument demonstrates that $m_1<0$ reduces the drop's propulsion velocity.

For ${\pe \geq 25}$ and ${m_1 = 2}$ the steady self-propulsion becomes unstable and steady axisymmetric pumping states are observed (i.e. with ${\textbf{U} = \textbf{0}}$ but non zero flow). The detailed dynamics of the spontaneous transition to steady pumping was analyzed extensively in Ref.~\cite{Morozov19b} for axisymmetric regimes, and we demonstrate below that this transition is not relevant in 3D as symmetric pumping states are unstable.

\subsection{Stability of the axisymmetric flow regimes}
\label{linear_stability}
\begin{figure*}
  \centering
  \raisebox{1.87in}{(a)}
  \includegraphics[scale=0.5]{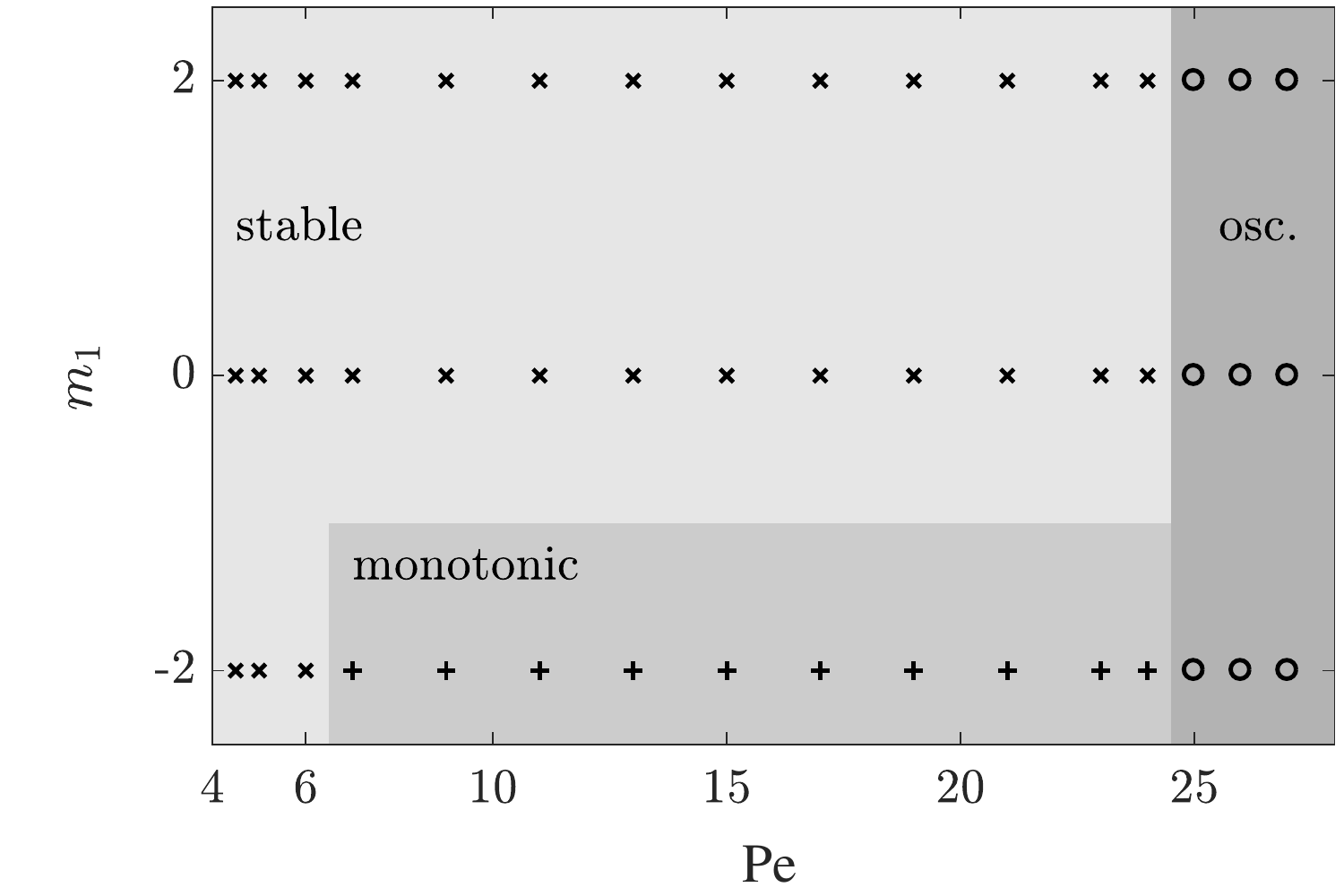}
  \quad \quad
  \raisebox{1.87in}{(b)}
  \includegraphics[scale=0.5]{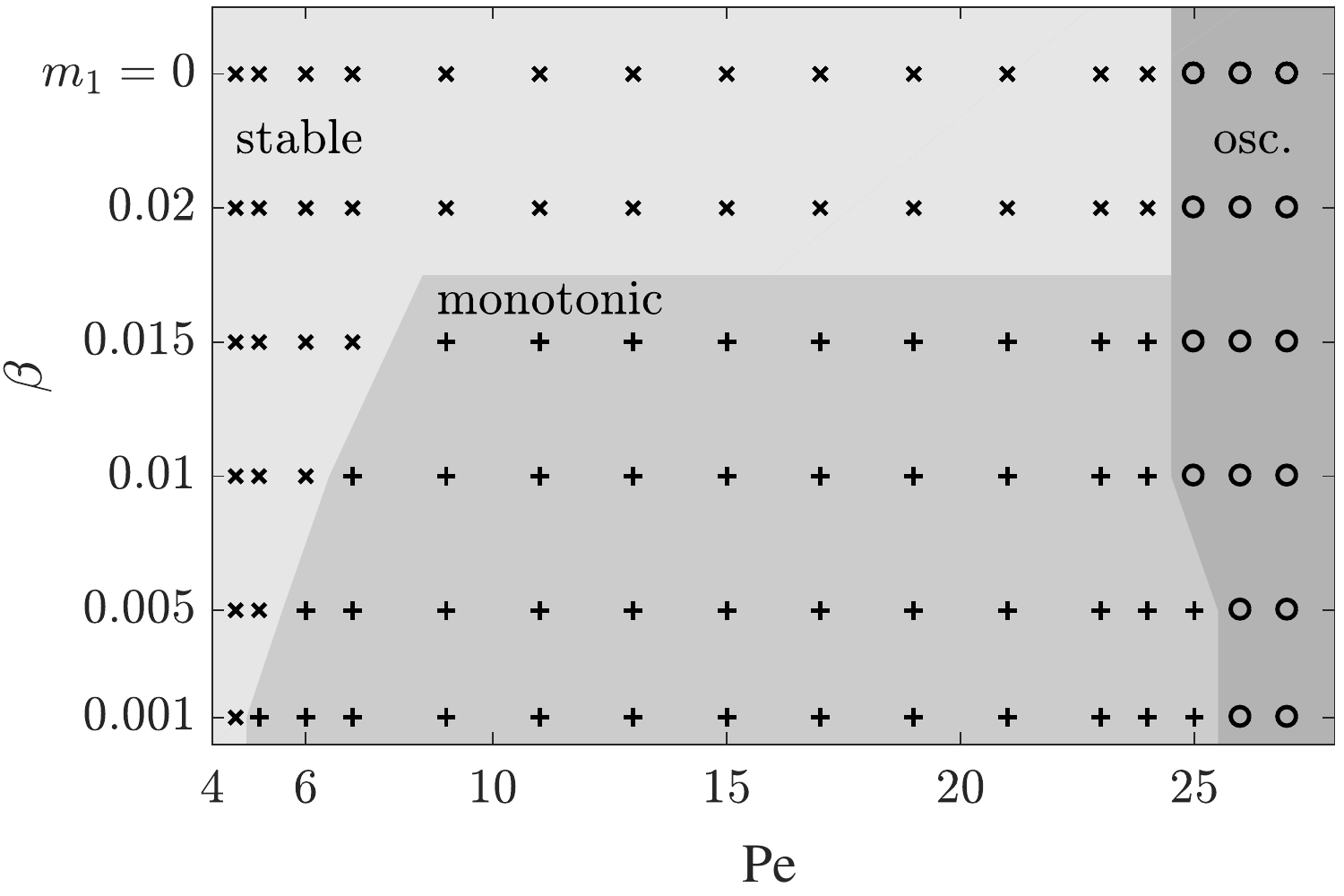}
  \caption{
    Stability maps of the axisymmetric self-propelling state with respect to 3D perturbations in the case (a) for ${\beta=0.01}$ and varying polarizability $m_1$ and P\'eclet number $\pe$, and (b) for $m_1=-2$ and varying retardation ratio $\beta$ and $\pe$. Here $\times$ denotes the stable steady state of axisymmetric propulsion, $+$ corresponds to a monotonic instability with respect to the perturbations of droplet self-propulsion direction, and $\circ$ marks an oscillatory instability that precedes the transition to chaos. In (b), results for ${m_1 = 0}$ (drop in isotropic state) are reported for reference.
  }
  \label{fig56}
\end{figure*}

Stability of the steady regimes shown in Fig.~\ref{fig3} with respect to infinitesimal non-axisymmetric perturbations is then analyzed (see Appendix~\ref{linear} for specifics on the numerical implementation), and its results are summarized in the stability maps shown in Fig.~\ref{fig56}. These results demonstrate that an active LC drop in nematic state with ${m_1 = -2}$ exhibits a new mode of instability that is not observed in the isotropic case (${m_1 = 0}$) and that emerges prior to the emergence of chaotic regimes (Figure~\ref{fig56}).

We explain the emergence of this new instability mode as follows. In the absence of nematic ordering, the problem formulated by Eqs.~\eqref{displacement}-\eqref{far} becomes isotropic. Without a preferred spatial direction, steady self-propulsion of isotropic drops in 3D must be neutrally stable with respect to a small change of propulsion direction (formally, perturbations in the form of small rotations of $\textbf{U}$ have a growth rate equal to zero). In contrast, in nematic state, any small change in the self-propulsion direction results in a misalignment between $\textbf{U}$ and $\textbf{q}$ that yields ${\dot{\textbf{q}} \neq 0}$ (i.e.~ perturbations in the propulsion direction have non-zero growth rates).

\blu{Recall that in active droplets self-propulsion velocity $\textbf{U}$} emerges from a self-generated concentration gradient; for ${m_1 < 0}$, the surface distribution of surfactants rotates $\textbf{q}$ away from the concentration gradient, Eq.~\eqref{omega}, thereby exacerbating any misalignment of $\textbf{U}$ and $\textbf{q}$, as illustrated in Fig.~\ref{fig7}a. Small rotations of the drop have a positive growth rate in this case, and the steady axisymmetric self-propulsion regime becomes unstable. The opposite is also true: when ${m_1 > 0}$, $\textbf{q}$ is rotated toward from the concentration gradient, thus damping out perturbations in the droplet polarization. 

The instability mechanism described above requires a persistence of the misalignment between the drop's propulsion velocity $\textbf{U}$ and its nematic polarization $\textbf{q}$. According to Eq.~\eqref{displacement}, this polarization relaxes toward the self-propulsion direction with a characteristic time ${\cal B}$. If ${\cal B}$ is too small, the alignment will be restored too quickly for the instability to develop, as shown in Fig.~\ref{fig56}b. In essence, we find that the instability requires a retardation ratio ${\beta \leq 0.015}$, that is, rearrangement of the director field within nematic drop must occur one order of magnitude slower than the time scale associated with the self-propelling flow, ${|\textbf{U}|R/{\cal V}}$.

When the P{\'e}clet number $\pe$ is sufficiently large, our linear analysis also predicts an oscillatory instability. \blu{By definition, oscillatory instability is a Hopf bifurcation, and, thus, features complex perturbation growthrates and complex eigenmodes as shown in Fig.~\ref{fig8}a.} Below we employ fully nonlinear simulations to demonstrate that this instability precedes the advection-driven transition to chaos.

\subsection{3D self-propulsion regimes}
\begin{figure*}
\centering
  \raisebox{1.75in}{\small(a)}
  \raisebox{0.08in}{
    \includegraphics[scale=0.45]{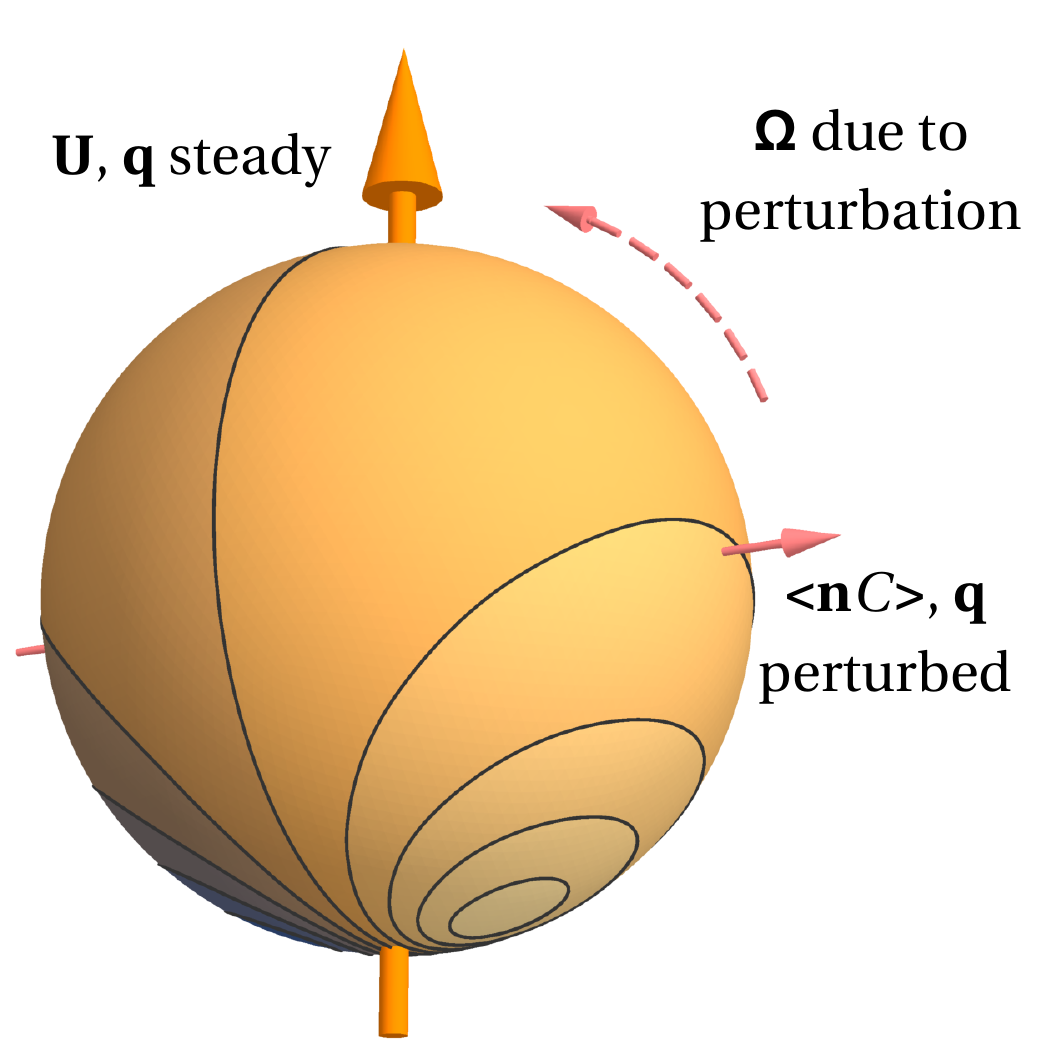}}
  \quad
  \includegraphics[scale=0.47]{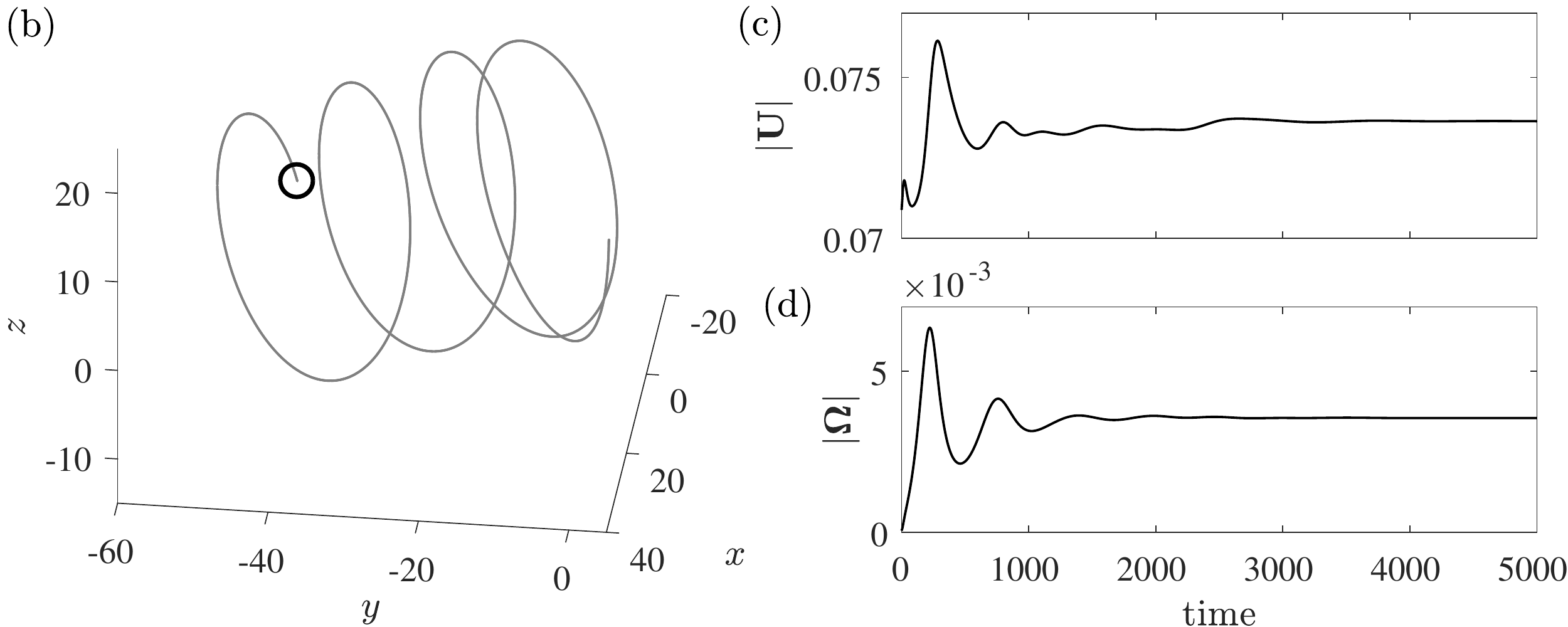}
  \caption{
    (a) Unstable eigenmode corresponding to the monotonic instability
    of an axisymmetric steady self-propulsion regime
    in the case of a nematic drop with
    ${\pe = 17}$, ${m_1 = -2}$, and ${\beta = 0.005}$.
    Contour plot represents surfactant concentration field
    at the surface of the drop, 
    where brighter color marks the regions of higher concentration.
    (b) Trajectory of an active nematic drop above the threshold 
    of the monotonic instability in the case of 
    ${\pe = 17}$, ${m_1 = -2}$, and ${\beta = 0.005}$.
    Evolution of drop translation $\textbf{U}$ 
    and rotation velocity $\boldsymbol \Omega$ 
    is shown in panels (c) and (d).
  }
  \label{fig7}
\end{figure*}

\begin{figure*}
\centering
  \raisebox{1.75in}{\small(a)}
   \raisebox{0.08in}{
    \includegraphics[scale=0.4]{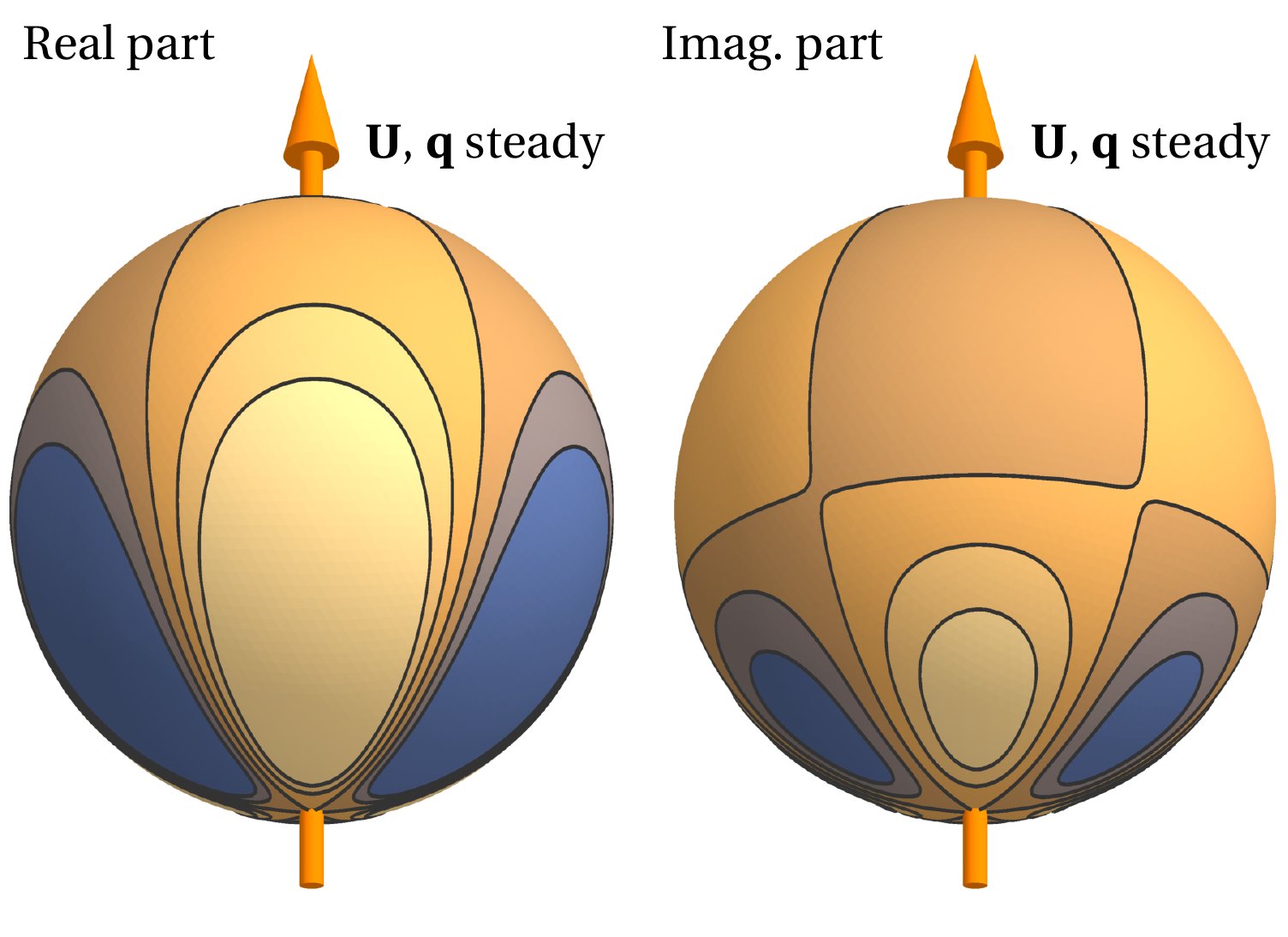}}
  \quad
  \includegraphics[scale=0.47]{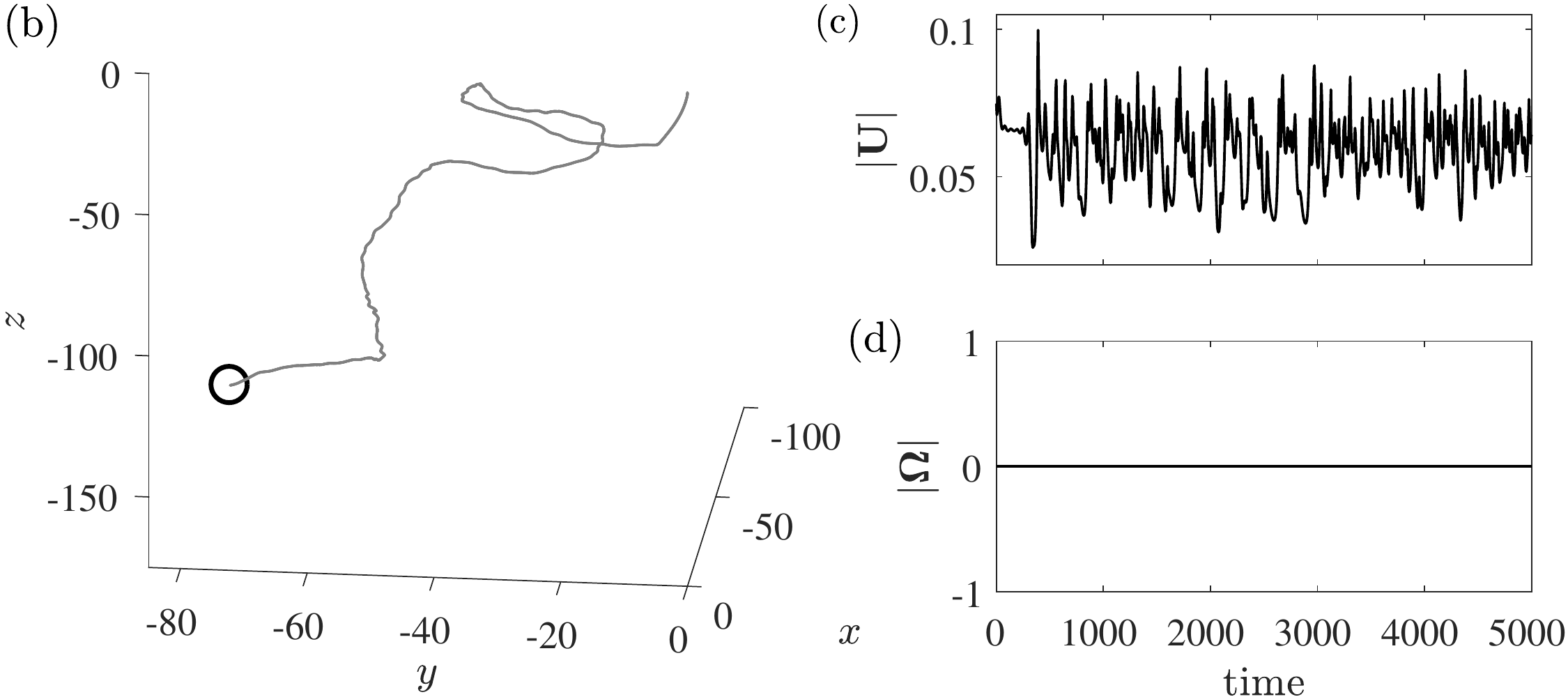}
  \caption{
    (a) Real (left) and imaginary part (right)
    of an eigenmode corresponding to the oscillatory instability
    of an axisymmetric steady self-propulsion regime
    in the case of an isotropic drop with
    ${\pe = 26}$ and ${m_1 = 0}$.
    Contour plot represents surfactant concentration field
    at the surface of the drop, 
    where brighter color marks the regions of higher concentration.
    (b) Trajectory of an isotropic drop above the threshold of 
    secondary oscillatory instability in the case of 
    ${\pe = 26}$ and ${m_1 = 0}$.
    Evolution of drop translation $\textbf{U}$ 
    and rotation velocity $\boldsymbol \Omega$ 
    is shown in panels (c) and (d).
  }
  \label{fig8}
\end{figure*}

We now explore the nonlinear regimes emerging above the thresholds of the secondary instabilities discussed in Sec.~\ref{linear_stability}. To this end, the full 3D evolution equations for $\textbf{q}$ and $C$ are solved numerically (see Appendix~\ref{method}) and the results are reported in Figs.~\ref{fig7} and~\ref{fig8}. Specifically, Figure~\ref{fig7} demonstrates that for ${\pe = 17}$, ${m_1 = -2}$, and ${\beta = 0.005}$, monotonic instability of the self-propulsion direction of nematic drops results in a helical motion of the drop. Interestingly, at the onset of instability, the translation velocity of the drop, $\textbf{U}$, is nearly orthogonal to the rotation velocity $\boldsymbol \Omega$ emerging from the perturbations, as shown in Fig.~\ref{fig7}a. 
Naturally, when ${\textbf{U} \perp \boldsymbol \Omega}$, drop trajectory should be a circle and, indeed, the trajectory obtained in simulations resembles a circle for ${t \lessapprox 1000}$ before evolving into a spiral trajectory in the fully nonlinear regime (Figure~\ref{fig7}b-d).

We further observe that the oscillatory instability setting in for high P{\'e}clet number, precedes the onset of chaos, as illustrated in Figure~\ref{fig8} for a drop in isotropic state (${m_1 = 0}$). \blu{It is easy to see that the eigenmode of oscillatory instability shown in Fig.~\ref{fig8}a possesses 3-fold rotational symmetry. This feature distinguishes this instability from the orientational instability illustrated in Fig.~\ref{fig7}. Indeed, the eigenmode of the orientational instability is asymmetric and, thus, is associated with rotation of the droplet, as we argue in the paragraph above. In contrast, symmetric perturbations, like the one shown in Fig.~\ref{fig8}a, are characterized by ${\left< \textbf{n} \left. C \right|_{r=1} \right> = \textbf{0}}$ and, therefore, may not impact droplet trajectory directly. Instead, the effect of this instability manifests itself in the droplet motion only via the nonlinear advective coupling of different eigenmodes. We further note that} oscillatory instability threshold remains roughly the same for both isotropic and nematic drops. Therefore, we attribute the instability as well as subsequent transition to chaos to the nonlinear effect of advection.

Advection-driven transition to chaos in isotropic active droplets was recently analyzed in the case of an axisymmetric flow, where chaotic self-propulsion trajectories emerge from an instability of a symmetric pumping state, characterized by ${\textbf{U} = \textbf{0}}$~\cite{Morozov19b}. In turn, the present simulations demonstrate that in 3D, the onset of chaos may be an instability of a self-propelling state with ${\textbf{U} \neq \textbf{0}}$. Note that in axisymmetric simulations chaotic motion of the drop is one order of magnitude slower than drop's steady self-propulsion~\cite{Morozov19b}. \blu{Indeed, in 3D numeral simulations and experiments~\cite{Suga18}, there is a continuum of possible directions for the drop to self-propel along, while in the axisymmetric case only two directions of motion are available. The number of allowed directions of motion does not matter in the steady self-propulsion regime, but is crucial for the chaotic motion, since the latter implies that the droplet may occasionally turn. It is easy to see that in order to turn in the axisymmetric case, droplet must first stop. That is, drop trajectory in phase space must pass near a fixed point solution corresponding to the motionless drop. Naturally, drop dynamics in vicinity of the fixed point must be slow, hence the significant decrease in the drop velocity, compared to the unrestricted 3D case.} In contrast, transition to chaotic motion in 3D is not associated with a significant decrease in the droplet's velocity (Figure~\ref{fig8}c), which is also consistent with experimental observations~\cite{Suga18}.

\section{Conclusions}
\label{discussion}
In this paper, we develop a minimal model for an active nematic liquid crystal (LC) drop suspended in the bulk of a surfactant solution in the case of homeotropic anchoring at the drop surface. This model is intrinsically isotropic, in the sense that there is no built-in asymmetry or chirality. Yet, we demonstrate that besides the symmetry-breaking transition to self-propulsion already established for active droplets, this model is also able to capture the spontaneous emergence of chirality and development of helical trajectories. \blu{Success of our approach is not accidental, as our model retains two key features of the transition from hedgehog to boojum defect observed in active LC microdrops. First, we note that in the course of this transition, nematic configuration within the droplet remains cylindrically symmetric. In this situation, orientation of the symmetry axis provides the most basic characteristic of the drop's nematic configuration and orientation of $\textbf{q}$ implements this characteristic in our model. Second, it is apparent that the flow can not push the boojum defect beyond the drop's surface. That is, there must be certain mechanism limiting the effect of the flow field on the nematic configuration. In our model this effect is naturally limited, since we postulate that in the steady state ${\left| \textbf{q} \right| \propto V}$, where $V$ is droplet self propulsion velocity that is saturated due to nonlinearity of surfactant advection, as shown in Fig.~\ref{fig3}. The main finding of the paper is that these two key features are enough to explain the emergence of helical trajectories in active LC microprops.}

Indeed, a novel mode of instability emerges due to the coupling of flow-induced nematic ordering and surfactant transport around a steadily self-propelling LC drop. This instability does not exist for isotropic active drops and results from the competition of two nematic reorientation mechanisms: the flow-induced nematic reordering (e.g. hedgehog defect displacement) and solid body rotation of the droplet. In particular, in the case when hedgehog defect displacement reduces mobility at the front of a moving drop (${m_1 < 0}$), self-propelling flow strives to align $\textbf{q}$ along the drop's velocity $\textbf{U}$, while any non-axisymmetric perturbation results in $\textbf{q}$ rotating away from the swimming direction (Figure~\ref{fig1}). Our linear stability analysis further reveals that the eigenmode of this novel instability features a drop's rotation velocity $\boldsymbol \Omega$ nearly orthogonal to $\textbf{U}$. The drop trajectory near the onset of instability should be close to a circle which is indeed confirmed in fully nonlinear simulations, before growing perturbations of $\textbf{U}$ become significant yielding ${\textbf{U} \not\perp \boldsymbol \Omega}$ and the drop trajectory becomes helical (Figure~\ref{fig7}).

Our nonlinear simulations also indicate that for sufficiently high P{\'e}clet numbers (slow diffusion), both isotropic and nematic drops exhibit transition to chaos and random trajectories (Figure~\ref{fig8}). The onset of chaos is preceded by an oscillatory instability characterized by highly symmetric eigenmodes, whose threshold is nearly insensitive to problem parameters besides $\pe$, hinting that surfactant advection is the main nonlinear mechanism enabling the transition to chaos in both isotropic and nematic droplets. Interestingly, and in contrast with the axisymmetric problem, 3D chaotic motion is not associated with a significant decrease in self-propulsion velocity.

The present \blu{minimal} model correctly captures the transitions between three main dynamical regimes recently observed in active LC drops, namely, self-propulsion with straight, helical, and random trajectories~\cite{Kruger16, Suga18} for increasing drop radius (i.e. increasing $\pe$). Emergence of helical trajectories is of particular importance, since it highlights how chirality may spontaneously arise in a polarizable physicochemical system due to the competition of reorientation mechanisms. The insight that emergence of chirality requires a certain degree of polarization is corroborated by experimental observations of cells that developed chiral properties while being exposed to a gradient of protein concentration~\cite{Chin18}. We expect that our theoretical arguments will facilitate the future modeling effort in physical chemistry and \blu{biofluidics} as well as motivate further use of active droplets as model objects.

In addition, our model brings important insight into the physicochemical phenomena at the interface of active drops: a faithful reproduction of the droplet dynamics can be achieved in the absence of the Marangoni effect which is commonly cited as the sole mechanism of active droplet mobility~\cite{Herminghaus14, Maass16}. Our results suggest therefore that diffusiophoresis in the form of slip velocity may be an important contributor to the hydrodynamics of surfactant-laden interfaces near and above critical micelle concentration.

\section*{Conflicts of interest}
There are no conflicts to declare.

\appendix
\section{Implementation of the numerical method}
\label{method}

\subsection{Real spherical harmonics}
\label{harmonics}
The numerical method used here is based on an expansion of the surfactant concentration in terms of real spherical harmonics $\bar{Y}_i^l( \mu, \phi )$ defined as,
\begin{equation}
  \label{real_Y}
  \bar{Y}_i^l( \mu, \phi ) \equiv \begin{cases}
    \alpha_i^l L_i^l( \mu ) \cos( l \phi ) \quad & l \geq 0\\
    \alpha_i^l L_i^{-l}( \mu ) \sin( -l \phi ) \quad & l < 0
  \end{cases}.
\end{equation}
Here $L_i^l$ are associated Legendre polynomials, ${\mu \equiv \cos \theta}$, and $\theta$ and $\phi$ are polar and azimuthal angles of the spherical coordinate system ${( r, \theta, \phi )}$, respectively, with $r=0$ corresponding to the droplet center. The normalization coefficients $\alpha_i^l$ are given by,
\begin{equation}
  \alpha_i^l \equiv \begin{cases}
    \sqrt{( 2 i + 1 )/( 4 \pi )} \quad & l = 0 \\
    \sqrt{( 2 i + 1 )( i - |l|)! / ( 2 \pi ( i + |l|)! )} \quad & l \neq 0
  \end{cases}.
\end{equation}

\subsection{3D Stokes flow around a sphere}
\label{stokes_flow}
Stokes flow around a spherical drop can be expressed in terms of vector spherical harmonics using the classical Lamb solution~\cite{Lamb45, Happel83, Leal07}. For a force-free squirming sphere, the flow in the co-moving reference frame moving writes~\cite{Pak14},
\begin{equation}
  \label{u_modes}
  \textbf{u}_i^l 
  = a_i^l(t) \big( f_i(r), g_i(r), g_i(r) \big) 
    \cdot \nabla \textbf{r} \bar{Y}_i^l
    + b_i^l(t) r^{-i-1} \nabla \times \textbf{r} \bar{Y}_i^l,
\end{equation}
where $a_i^l(t)$ and $b_i^l(t)$ are unknown time-dependent amplitudes, and $f_i(r)$ and $g_i(r)$ are defined as,
\begin{align}
  & f_i(r) = \begin{cases}
    2 \sqrt{\pi/3} ( 1 - r^{-3} ) \quad & i = 1 \\
    ( i + 1 ) ( r^{-i} - r^{-i-2} ) \quad & i > 1
  \end{cases}, \\
  & g_i(r) = \begin{cases}
    \sqrt{\pi/3} ( 2 + r^{-3} ) \quad & i = 1 \\
    -( i - 2 ) r^{-i} / i + r^{-i-2} \quad & i > 1
  \end{cases}.
\end{align}
We note that in Cartesian coordinates drop translation velocity is given by ${\textbf{U} = -\left( a_1^1, a_1^{-1}, a_1^0 \right)}$.

\subsection{Solving for the nonlinear dynamics}
\label{numerics}
Our approach to the problem formulated by Eqs.~\eqref{displacement}-\eqref{far} is based on the method developed by Michelin and Lauga~\cite{Michelin11}. Specifically, we approximate the flow field and concentration distribution using a finite number of real spherical harmonics~\eqref{real_Y},
\begin{align}
  \label{u_exp_num}
  & \textbf{u}( t, r, \mu, \phi ) 
    \approx \sum\limits_{i=0}^N \sum\limits_{l=-i}^i 
      \textbf{u}_i^l ( t, r, \mu, \phi ) , \\
  \label{c_exp_num}
  & C( t, r, \mu, \phi ) \approx \sum\limits_{i=0}^N \sum\limits_{l=-i}^i 
    C_i^l( t, r ) \bar{Y}_i^l( \mu, \phi ).
\end{align}

Substitution of Eqs.~\eqref{u_exp_num}--\eqref{c_exp_num} into the dimensionless form of the boundary condition, Eq.~\eqref{slip}, and projection of the result onto the two families of orthogonal modes yields, 
\begin{align}
  \label{a_sol}
  & a_i^l = A_i \left. C_i^l \right|_{r=1}
    + m_1 B_i \sum\limits_{j=1}^N \sum\limits_{m=-j}^j \sum\limits_{n=-1}^1
        J_{ji1}^{mln} q^n \left. C_j^m \right|_{r=1}, \\
  \label{b_sol}
  & b_i^l = \dfrac{2 m_1}{i (i+1)} \sqrt{\dfrac{\pi}{3}}
    \sum\limits_{j=1}^N \sum\limits_{m=-j}^j \sum\limits_{n=-1}^1
        K_{ji1}^{mln} q^n \left. C_j^m \right|_{r=1}, \\
  \label{omega_sol}
  & \qquad \quad \boldsymbol \Omega = \frac{m_1}{4} \sqrt{\dfrac{3}{\pi}}
    \textbf{q} \times \left. \left( C_1^1, C_1^{-1}, C_1^0 \right) \right|_{r=1},
\end{align}
where 
\begin{align}
  & A_i = \begin{cases}
    -1/\sqrt{3 \pi} \; \; & i = 1 \\
    -i/2 \; \; & i > 1
  \end{cases}, \\
  & B_i = \begin{cases}
    -1/3 \; \; & i = 1 \\
    -\sqrt{\pi/3}/(i+1) \; \; & i > 1
  \end{cases}, \\
  \label{j_int}
  J_{ijk}^{lmn} \equiv & \int\limits_{-1}^1 d\mu \int\limits_0^{2\pi} d\phi
    \left( 1 - \mu^2 \right) \left( \partial_\mu \bar{Y}_i^l \right) 
      \left( \partial_\mu \bar{Y}_j^m \right) \bar{Y}_k^n \nonumber \\
    & + \int\limits_{-1}^1 d\mu \int\limits_0^{2\pi} d\phi
      \frac{ \left( \partial_\phi \bar{Y}_i^l \right) 
        \left( \partial_\phi \bar{Y}_j^m \right)
          \bar{Y}_k^n(\mu) }{\left( 1 - \mu^2 \right)}, \\
  \label{k_int}
  K_{ijk}^{lmn} \equiv & \int\limits_{-1}^1 d\mu \int\limits_0^{2\pi} d\phi
    \left( \partial_\mu \bar{Y}_i^l \right) 
      \left( \partial_\phi \bar{Y}_j^m \right) \bar{Y}_k^n \nonumber \\
    & - \int\limits_{-1}^1 d\mu \int\limits_0^{2\pi} d\phi
      \left( \partial_\phi \bar{Y}_i^l \right) 
        \left( \partial_\mu \bar{Y}_j^m \right)
          \bar{Y}_k^n(\mu),
\end{align}
and the components of the effective polarization vector $\textbf{q}$ are arranged to match the modal expansion, namely, ${\left( q^1, q^{-1}, q^0 \right) \equiv \textbf{q}}$.

Finally, substitution of Eqs.~\eqref{u_exp_num}-\eqref{c_exp_num} into the dimensionless form of the advection-diffusion equation, Eq.~\eqref{eqs_ad}, and projection of the result onto $\bar{Y}_k^n$ yields a set of coupled nonlinear differential equations for $C_k^n( t, r )$, namely,
\begin{multline}
\label{ckn_evo}
  \partial_t C_k^n
  = \frac{1}{\pe} \left( 
    \partial_{rr} C_k^n + 2 \frac{\partial_r C_k^n}{r} 
    - \frac{k \left( k + 1 \right) C_k^n}{r^2} \right)
    \\ - \sum\limits_{i=1}^N \sum\limits_{l=-i}^i 
    \sum\limits_{j=0}^N \sum\limits_{m=-j}^j \bigg(
      a_i^l \bigg[ I_{ijk}^{lmn} f_i \partial_r C_j^m
        + \frac{J_{ijk}^{lmn} g_i C_j^m}{r} \bigg]
    \\ + \frac{b_i^l K_{ijk}^{lmn} C_j^m}{r^{i+2}} \bigg),
\end{multline}
where
\begin{equation}
  \label{i_int}
  I_{ijk}^{lmn} \equiv \int\limits_{-1}^1 d\mu \int\limits_0^{2\pi} d\phi
    \, \bar{Y}_i^l \bar{Y}_j^m \bar{Y}_k^n.
\end{equation}

Together with dimensionless form of Eq.~\eqref{displacement}, Eqs.~\eqref{ckn_evo} constitute a closed problem describing the evolution of the concentration distribution around a self-polarizing active drop. Similarly to the axisymmetric case considered in Refs.~\cite{Michelin13a,Michelin14,Morozov19b}, we solve Eq.~\eqref{ckn_evo} numerically, using an exponentially-stretched radial grid (i.e. $r = e^{\xi^3 - 1}$, with  evenly-spaced $\xi$). A Crank-Nicholson scheme is used for the diffusive term and an explicit time-stepping scheme for the advective term, while the evolution equation for the self-polarization vector is integrated explicitly. The convergence of the modal approximations, Eqs.~\eqref{u_exp_num}-\eqref{c_exp_num}, is ensured by repeating all of the computations for $N = 15$ (225 modes in total) and $18$ (324 modes).
Spatial grids with $60$ and $120$ nodes and the time step of $0.05$ and $0.02$, respectively, were used to obtain the results of this paper.

\subsection{Linear analysis}
\label{linear}
In the course of analysis we demonstrate that evolution equations, Eqs.~\eqref{displacement} and~\eqref{ckn_evo}, admit a branch of axisymmetric steady solutions, ${\widehat{\textbf{q}} = \left( 0, 0, \widehat{q}^0 \right)}$ and ${\widehat{C}_k^0 (r)}$, corresponding to a drop that self-propels along a straight line. To access the stability of this self-propelling state, the dimensionless form of Eq.~\eqref{displacement} and Eq.~\eqref{ckn_evo} are linearized,
\begin{align}
  \textbf{q}(t) & = \left( 0, 0, \widehat{q}^0 \right)
    + \widetilde{\textbf{q}} e^{\lambda t}, \\
  C_k^n(t,r) & = \widehat{C}_k^0(r) + \widetilde{C}_k^n(r) e^{\lambda t},
\end{align}
where $\lambda$ is the growth rate of an infinitesimal perturbation denoted by tilde.

The linearized equations include various products of the steady quantities, $\widehat{q}^0$, $\widehat{C}_k^0$, and perturbations, $\widetilde{\textbf{q}}$, $\widetilde{C}_k^n$, that can be simplified using the following identities,
\begin{align}
  \label{i_simp}
  I_{ijk}^{0mn} = I_{ijk}^{0nn} \delta_{mn}, & \quad
  I_{ijk}^{l0n} = I_{ijk}^{n0n} \delta_{ln}, \\
  \label{j_simp}
  J_{ijk}^{0mn} = J_{ijk}^{0nn} \delta_{mn}, & \quad
  J_{ijk}^{l0n} = J_{ijk}^{n0n} \delta_{ln}, \\
  \label{k_simp}
  K_{ijk}^{0mn} = K_{ijk}^{0-nn} \delta_{m-n}, & \quad
  K_{ijk}^{l0n} = K_{ijk}^{-n0n} \delta_{l-n}.
\end{align}
Identities~\eqref{i_simp}-\eqref{k_simp} imply that linearized equations~\eqref{displacement} and~\eqref{ckn_evo} reduce to a set of ${N + 1}$ independent linear problems, where each problem implements the stability of the steady state $\widehat{C}_k^0(r)$ with respect to the family of perturbations $\widetilde{C}_k^n(r)$ with a given azimuthal order (i.e. a given value of $|n|$). The Jacobian of the right-hand sides of Eqs.~\eqref{displacement} and~\eqref{ckn_evo} is constructed numerically for a given $n$ in order to obtain its eigenvalues and the growth rate of the perturbations.

\section*{Acknowledgements}
This project has received funding from the European Research Council (ERC) under the European Union's Horizon 2020 research and innovation programme (grant agreement No 714027 to SM).

\balance

\bibliography{refs} 

\end{document}